\documentclass[twocolumn]{aastex63}

\definecolor{red}{rgb}{0.8,0.0,0.0}
\definecolor{blue}{rgb}{0.0,0.0,0.8}
\definecolor{green}{rgb}{0.0,0.5,0.0}
\usepackage{comment}

\shorttitle{Neutral metals in the atmosphere of HD149026b}
\shortauthors{Ishizuka et al.}

\graphicspath{{./}{figures/}}

\begin{document}

\title{Neutral metals in the atmosphere of HD149026b}
\correspondingauthor{Masato Ishizuka}
\email{ishizuka@astron.s.u-tokyo.ac.jp}

\author[0000-0003-1906-4525]{Masato Ishizuka}
\affiliation{Department of Astronomy, The University of Tokyo, 7-3-1 Hongo, Bunkyo-ku, Tokyo 113-0033, Japan}

\author[0000-0003-3309-9134]{Hajime Kawahara}
\affiliation{Department of Earth and Planetary Science, The University of Tokyo, 7-3-1 Hongo, Bunkyo-ku, Tokyo 113-0033, Japan}
\affiliation{Research Center for the Early Universe, 
School of Science, The University of Tokyo, Tokyo 113-0033, Japan}

\author[0000-0003-4698-6285]{Stevanus K. Nugroho}
\affiliation{School of Mathematics and Physics, Queen's University Belfast, University Road, Belfast, BT7 1NN, United Kingdom}
\affiliation{National Astronomical Observatory of Japan, NINS, 2-21-1 Osawa, Mitaka, Tokyo 181-8588, Japan}
\affiliation{Astrobiology Center, NINS, 2-21-1 Osawa, Mitaka, Tokyo 181-8588, Japan}

\author[0000-0003-3800-7518]{Yui Kawashima}
\affiliation{Cluster for Pioneering Research, RIKEN, 2-1 Hirosawa, Wako, Saitama 351-0198, Japan}
\affiliation{SRON Netherlands Institute for Space Research, Sorbonnelaan 2, 3584 CA Utrecht, The Netherlands}

\author[0000-0003-3618-7535]{Teruyuki Hirano}
\affiliation{Department of Earth and Planetary Sciences, Tokyo Institute of Technology, 2-12-1 Ookayama, Meguro-ku, Tokyo 152-8551, Japan}

\author[0000-0002-6510-0681]{Motohide Tamura}
\affiliation{Department of Astronomy, The University of Tokyo, 7-3-1 Hongo, Bunkyo-ku, Tokyo 113-0033, Japan}
\affiliation{National Astronomical Observatory of Japan, NINS, 2-21-1 Osawa, Mitaka, Tokyo 181-8588, Japan}
\affiliation{Astrobiology Center, NINS, 2-21-1 Osawa, Mitaka, Tokyo 181-8588, Japan}


\begin{abstract}
Recent progress in high-dispersion spectroscopy has revealed the presence of vaporized heavy metals and ions in the atmosphere of hot Jupiters whose dayside temperature is larger than 2000 K, categorized as ultra hot Jupiters (UHJs). Using the archival data of high resolution transmission spectroscopy obtained with the Subaru telescope, we searched for neutral metals in HD149026b, a hot Jupiter cooler than UHJs. By removing stellar and telluric absorption and using a cross-correlation technique, we report tentative detection of neutral titanium with 4.4 $\sigma$ and a marginal signal of neutral iron with 2.8 $\sigma$ in the atmosphere. 
This is the first detection of neutral titanium in an exoplanetary atmosphere. In this temperature range, titanium tends to form titanium oxide (TiO). The fact that we did not detect any signal from TiO suggests that the C/O ratio in the atmosphere is higher than the solar value. The detection of metals in the atmosphere of hot Jupiters cooler than UHJs will be useful for understanding the atmospheric structure and formation history of hot Jupiters. 

\end{abstract}

\keywords{Exoplanet atmospheres, High resolution spectroscopy, Transmission spectroscopy, Hot Jupiters}
\section{Introduction}

Since the first detection of 51 Peg b by \citet{MayorQueloz1995}, the extreme environments of the planetary atmospheres of hot Jupiters have been extensively studied. 
Strong stellar irradiation leads to high atmospheric temperatures in these planets, and tidal locking of the orbit results in extreme atmospheric conditions such as large temperature discrepancies between the dayside and nightside, and equatorial jets \citep[e.g.][]{Louden_Wheatley2015}. 
 \\
Observations of the atmospheres of hot Jupiters are important for understanding their unique properties.
For example, one of the most important issues for hot Jupiters is how they are formed. 
\textit{In-situ} formation has been considered to be unlikely for hot Jupiters, because a large enhancement of the solid surface density is required to grow gas-giant planets.
However, recent studies have shown several scenarios for \textit{in-situ} formation of gas giants (e.g., \citealt{Batygin2016}, \citealt{Boley2016A}). Another possible scenario is planet migration. In general, there are two hypotheses for migration of hot Jupiters: disk-interaction during their formation (e.g., \citealt{Tanaka2002}) and dynamical scattering by other planets (e.g., \citealt{Rasio1996}). Because the atomic and molecular species in the planetary atmosphere reflects the abundance of the chemical elements and environment of the place where they grew, the atmospheric composition is an important clue for investigating these scenarios (e.g., \citealt{Madhusudhan2014}).

\begin{table*}
  \centering
  \begin{tabular}{c|c|c|c}
    Planet &\begin{tabular}{c}  Dayside\\temperature (K) \end{tabular}& \begin{tabular}{c}  Detected heavy metals \\ and molecules \end{tabular} & Reference  \\ \hline\hline
    KELT-9b & $\sim$5000 & Mg I, Fe I, Fe II, Ti II, Cr I, Cr II &   a,b,c  \\
            &       & Sc II, Y II, Co I, Sr II, Ca I, Ca II &   \\
    WASP-33b & $\sim$3300 & Ca II, Fe I, TiO & c,d,e \\
    KELT-20b & $\sim$2900 & Fe I, Fe II, Ca II & f,g,h \\
    WASP-121b & $\sim$2800 & Fe I, Fe II, Mg II, H$_2$O  & i,j,k\\
    WASP-76b & $\sim$2700 & Fe I & l \\
    HD149026b & $\sim$2100 & Fe I , Ti I & This paper 
    
  \end{tabular}
  \caption{Detected heavy metals and molecules in UHJ atmospheres. To calculate the dayside temperatures, we assumed instantaneous re-radiation (no day--night heat distribution) with Bond albedo = 0.1.\\
  a: \citet{Hoeij2018}, b: \citet{Hoeij2019}, c: \citet{Yan2019}, d: \citet{Nugroho2017}, e: \citet{Nugroho2020_wp33fe}, f: \citet{Casasayas2018}, g: \citet{Casasayas2019}, h: \citet{Nugroho2020}, i: \citet{Evans2017}, j: \citet{Sing2019}, k \citet{Gibson2020}, l: \citet{Ehrenreich2020}}
  \label{tb:UHJ}
\end{table*}

Transmission and dayside spectroscopy from space have revealed diverse atmospheric properties of exoplanets (e.g., \citealt{Charbonneau2008}, \citealt{Pont2009}). 
High dispersion spectroscopy from the ground with a cross-correlation technique has also been recognized as a powerful tool to characterize exoplanetary atmospheres. 
Absorption/emission lines of atoms and molecules are resolved in high-dispersion spectroscopy, allowing their robust detection. 
This method has successfully been used to detect molecules such as CO, ${\rm H_2 O}$, TiO, and HCN  (e.g., \citealt{ Snellen2010, Brogi2012, Birkby2013,Nugroho2017,Hawker2018}).
If the atmospheric temperature is sufficiently high, the exoplanetary atmosphere contains vaporized metals (e.g., \citealt{Kitzmann2018}).
Hot Jupiters with dayside temperatures $\geq$ 2000 K have been called ultra hot Jupiters (UHJs) (\citealt{Fortney2008}, \citealt{Parmentier2018}),
and recent studies of UHJs by high-resolution transmission spectroscopy have reported vaporized metals in the atmosphere, as summarized in Table 1. 

The high temperature of UHJs should lead to their atmospheres being cloud-free.
Clouds in the planetary atmosphere make the transmission spectrum featureless and make it difficult to detect chemical species in the atmosphere. 
The high temperature should also lead to a large amount of emission and their atmospheres being inflated, thus UHJs' signal is large in both emission spectroscopy and transmission spectroscopy. 
Therefore, UHJs are suitable targets for investigating the chemical compositions of planetary atmospheres.\\
The UHJs that have been well studied to date have dayside temperatures $\geq$ 2500 K, while the atmospheric properties of UHJs cooler than 2500 K have not been well investigated. 
In this paper, we report the results of the high dispersion transmission spectroscopy of HD149026b, whose dayside temperature is $\sim$ 2100 K.
HD 149026 is a subgiant G0 IV star with a radius of ${\rm 1.4\ R_{Sun}}$ and a mass of ${\rm 1.4\ M_{Sun}}$. \citet{Sato2005} discovered a Saturn-sized planet around the star by transit and radial velocity measurements. The high density suggests a large and massive core with a mass of  ${\rm \sim 67\ M_{\oplus}}$. The process of its formation is still in debate. 
The dayside temperature of HD149026b is high enough for it to have vaporized metals in the atmosphere, though it is cooler than the other UHJs well studied so far.

We describe the observations and data reduction in Section 2. The cross-correlation technique we applied is explained in Section 3. In Section 4, we present the results of the analysis.  Section 5 is devoted to discussions and a summary of the paper.

\begin{table}
  \centering
  \begin{tabular}{p{40mm}c}
    Parameter & Value  \\ \hline\hline 
    {\bf HD149026} &    \\
    Radius (${\rm R_\odot}$)&  1.41$\pm ^{0.03}_{0.03} $ $^a$   \\
    Mass (${\rm M_\odot}$) & 1.42$\pm ^{0.33}_{0.33}$ $^a$  \\
    ${\rm T_{eff}}$ (K)& 6179$\pm ^{15}_{15}$ $^a$  \\
    log g & 4.37$\pm ^{0.04}_{0.04}$ $^a$ \\
    Spectral type & G0I\hspace{-.1em}V $^c$ \\
    v$\sin{i}$ (${\rm km \ s^{-1}}$) & 6.0$\pm ^{0.5}_{0.5}$ $^c$ \\
    ${\rm[Fe/H]}$ & 0.36$\pm ^{0.05}_{0.05}$ $^c$ \\ \hline
    {\bf HD149026b} & \\
    Radius (${\rm R_{Jup}}$)&  0.74$\pm ^{0.02}_{0.02}$ $^a$  \\
    Mass (${\rm M_{Jup}}$) & 0.38$\pm ^{0.014}_{0.012}$ $^a$ \\
    $T_0$ (BJD) &  2454597.70713$\pm ^{0.00016}_{0.00016}$ $^b$\\
    Period (days) & 2.8758916$\pm ^{0.000002}_{0.000002}$ $^b$\\
    Inclination (deg) &  84.55$\pm ^{0.58}_{0.58}$ $^a$ \\
    $K_p$ (${\rm km \ s^{-1}}$) & 169$\pm ^{20}_{20}$

  \end{tabular}
  \caption{Basic parameters of HD149026 and HD149026b. $K_p$ means the radial velocity semi-amplitude.\\
  $a$: \citet{Stassun2017}, $b$: \citet{Bonomo2017}}, $c$: \citet{Sato2005}
  \label{tb:Param}
\end{table}

\section{Observations and data reduction}

\subsection{Subaru observation}
 We analyzed the archival data for transmission spectroscopy of HD 149026, which were taken from the SMOKA system \citep{Baba2002}. The data were obtained on the night of May 11th, 2009, with the High Dispersion Spectrograph (HDS, \citealt{Noguchi2002}) on the Subaru telescope. A total of 41 frames were observed using the standard Ra (StdRa) mode with no iodine cell. The exposure time per frame ranged from 480 to 720 seconds and the mean was $\sim$ 550 seconds. 
Total exposure was 22680 seconds, and exposure during the transit was 10320 seconds.
 The slit width was 0.2 mm, which yields a wavelength resolution of R {\rm $\sim$} 90000. The spectra were taken by two (blue and red) CCDs, which contained 26 and 17 orders and covered the wavelength of 4923--6227 \AA \ and 6340--7660 \AA \, respectively.

\subsection{Standard reduction}
We conducted a standard reduction of the data by IRAF tools and {\sl hdsql}, provided by the team of HDS\footnote{https://www.naoj.org/Observing/Instruments/HDS/hdsql-e.html}. First, we performed an overscan correction,  removed the bias, and converted the analog digital unit to electron numbers.
We made mask frames of the bad columns for each CCD. These processes were performed by {\sl hdsql}. Scattered light was then subtracted by the IRAF task {\sl apscatter}. Using 50 flat frames, we generated a median flat frame with a non-linearity correction (\citealt{Tajitsu2010}). To remove the fringe pattern on the spectrum, we normalized the flat frame by {\sl apnormalize} and divided the object frames by the normalized flat frame. Then we extracted a 1-D spectrum by {\sl apall}. The grid of the wavelength was derived by the IRAF task {\sl ecidentify} using thorium--argon arc lump spectra, which were taken at the beginning and end of the night. We assigned the wavelength grid to each object frame by {\sl refspectra} and {\sl dispcor}.  We conducted these reductions to the median flat frame as well as the object frames to estimate the blaze function of each order. We divided the object spectra by the estimated blaze function using {\sl sarith}. The spectrum contains several regions affected by bad columns that were not completely removed by the reduction with the CL script. Because these bad regions can be problematic for analysis, we identified the bad regions by visual inspection and masked them.

\subsection{Correction for variations of blaze function}
\begin{figure*} 
\begin{center}
\includegraphics[width=17cm]{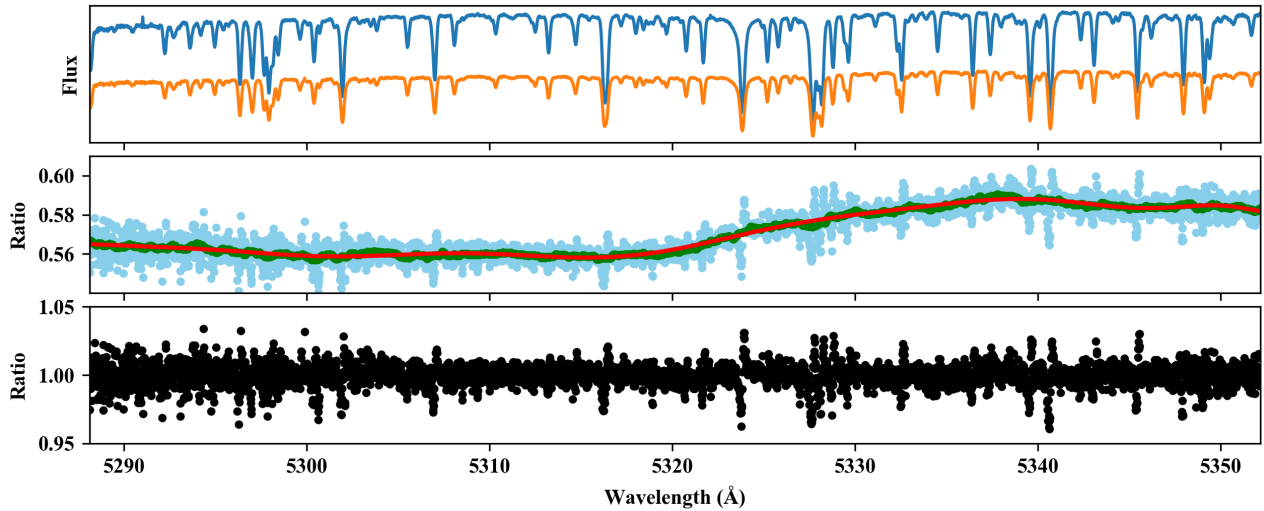}
\caption{Correction for blaze function variation.
The top panel shows the spectrum of the first frame and the reference frame (orange: 1st frame, blue: reference frame) of echelle order 112. The flux difference of these frames was mainly due to the difference in exposure time (480 seconds and 720 seconds, respectively). The middle panel shows the flux ratio for these two spectra. The light blue dots are the raw ratio and the green dots are after 3-sigma clipping and smoothing. 
The red line is a fit with the Chebyshev 15th order function. We used this fitted ratio as a compensator for the blaze function variation. The bottom panel shows the flux ratio for the corrected 1st frame and the reference frame.}
\label{fig:blcor}
\end{center}
\end{figure*}

  As described in previous research, the blaze function of HDS varies during an observation (e.g., \citealt{Winn2004}, \citealt{Narita2005}). To correct for this variation, we conducted the following procedure. \\
  First, we divided the spectrum of each frame by that of a reference frame. We used the 21st frame as the reference frame because its signal to noise ratio (S/N) was the highest. We took the raw ratio of each spectrum and the reference spectrum, and its change during the observation (Figure \ref{fig:blcor}). Then we performed a 3-sigma clipping and smoothing of the ratio to remove the effects of residual absorption lines. After that, we fitted the raw ratio with the Chebyshev 15th order function. The fitted ratio determined in this way represents the variation of the blaze function for the frame compared with the reference frame. We multiplied the reference spectrum by the raw ratio and divided it by the fitted ratio, which was a spectrum with the same blaze function as the reference spectrum.  Finally, we derived the continuum spectrum of the reference spectrum and normalized it by the IRAF task {\sl continuum}. Because the spectra of all the frames have the same blaze function, we normalized the spectra of the other frames by this continuum spectrum.

\subsection{Transmission spectrum matrix}
As described in section 2.3, we obtained the normalized raw spectra $S_\mathrm{raw}$ for 41 frames. 
For further analysis, we converted $S_\mathrm{raw}$ to the transmission spectrum matrix in both the Earth and stellar rest frames with common wavelength grids (denoted by $F_\mathrm{Earth}$ and $F_\mathrm{stellar}$, respectively).

\begin{figure}[ht] 
\begin{center}
\includegraphics[width=8cm]{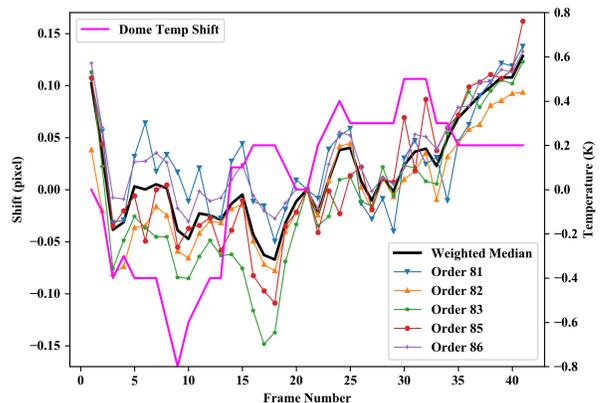}
\caption{Relative shift owing to instrumental variation. The black solid line is the weighted median shift of the five orders where a sufficient telluric line exists. The magenta line shows the variations of the dome temperature. The relative shift is correlated with the temperature variation in the dome of Subaru. \label{fig:shift}}
\end{center}
\end{figure}

Instrumental instability during the observation causes a time varying shift of the spectrum on the detector. We used the telluric absorption lines in $S_{raw}$ to measure the instrumental shift relative to the reference frame. The shift was estimated from the peak of the cross-correlation between $S_\mathrm{raw}$ and the model telluric transmission spectrum\footnote{https://www.eso.org/observing/etc/skycalc/skycalc.htm} for each echelle order \footnote{ We found that only five orders in the red CCD have sufficiently strong and clear telluric absorption lines. We used these orders to measure the instrumental shift. The weighted median of the central variations of these orders was used.}.  The cross-correlation function (CCF) was computed  from the velocity difference of $-$20 ${\rm km \ s^{-1}}$ to 20 ${\rm km \ s^{-1}}$ with a 0.1 ${\rm km \ s^{-1}}$ step. The peak of the CCF was derived by fitting a Gaussian to the CCF. The instrumental shift derived by the above procedure is shown in Figure \ref{fig:shift}. The shift correlates with the dome temperature. We shifted $S_\mathrm{raw}$ with spline interpolation. We conducted a 5-sigma clipping to remove bad pixels and cosmic rays and obtained $F_\mathrm{Earth}$.

Then, we generated $F_\mathrm{stellar}$ by a cross-correlation analysis of $F_\mathrm{Earth}$ with the theoretical stellar spectrum. The detailed procedure is as follows. We measured the apparent radial velocity of the star in each frame by the CCF with the stellar model spectrum (\citealt{Coelho2005}). We fitted a Gaussian to the CCF and used the center position of the best-fit Gaussian as the stellar radial velocity in each frame. The apparent radial velocity variation is {\rm $\sim$} 600 ${\rm m \ s^{-1}}$ between the first and last frames, which is consistent with the barycentric radial velocity variation of the Earth toward HD149026. Then we shifted the spectra according to the stellar radial velocity of each frame. We note that this correction includes both the barycentric velocity variation and the stellar Rossiter--McLaughlin (RM) effect. 

We also obtained the systemic radial velocity in this reduction. The systemic radial velocity used throughout this paper is $-$17.92 ${\rm \pm 0.04}$ ${\rm km \  s^{-1}}$.

\subsection{Removal of stellar spectrum and telluric absorption}

\begin{figure*} 
\begin{center}
\includegraphics[width=18cm]{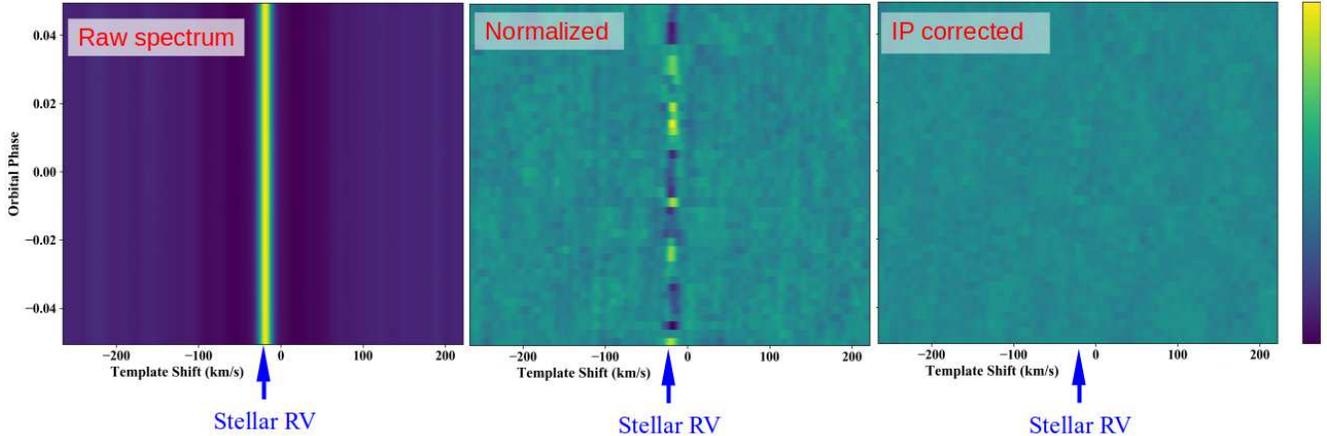}
\caption{CCF matrix of the model stellar spectrum in each reduction step described in section 2.4. The left figure is $C^{st}_\mathrm{stellar}$, namely, the CCF matrix of the transmission spectrum matrix $F_\mathrm{stellar}$ and the stellar model spectrum. The middle figure shows $C^{st}_\mathrm{norm}$, the CCF matrix of $F_\mathrm{stellar}$ normalized by out-transit spectra. The value of $C^{st}_\mathrm{norm}$ at the stellar radial velocity is shown in Figure \ref{fig:noisecorr}. The right figure shows $C^{st}_\mathrm{IP}$, the CCF matrix of $F_\mathrm{stellar}$ with IP correction. Since this is a cross-correlation, the scale is arbitrary, but the middle and right figure are shown with the same color scale for comparison.\label{fig:ccfresidual}}
\end{center}
\end{figure*}

\begin{figure*} 
\begin{center}
\includegraphics[width=18cm]{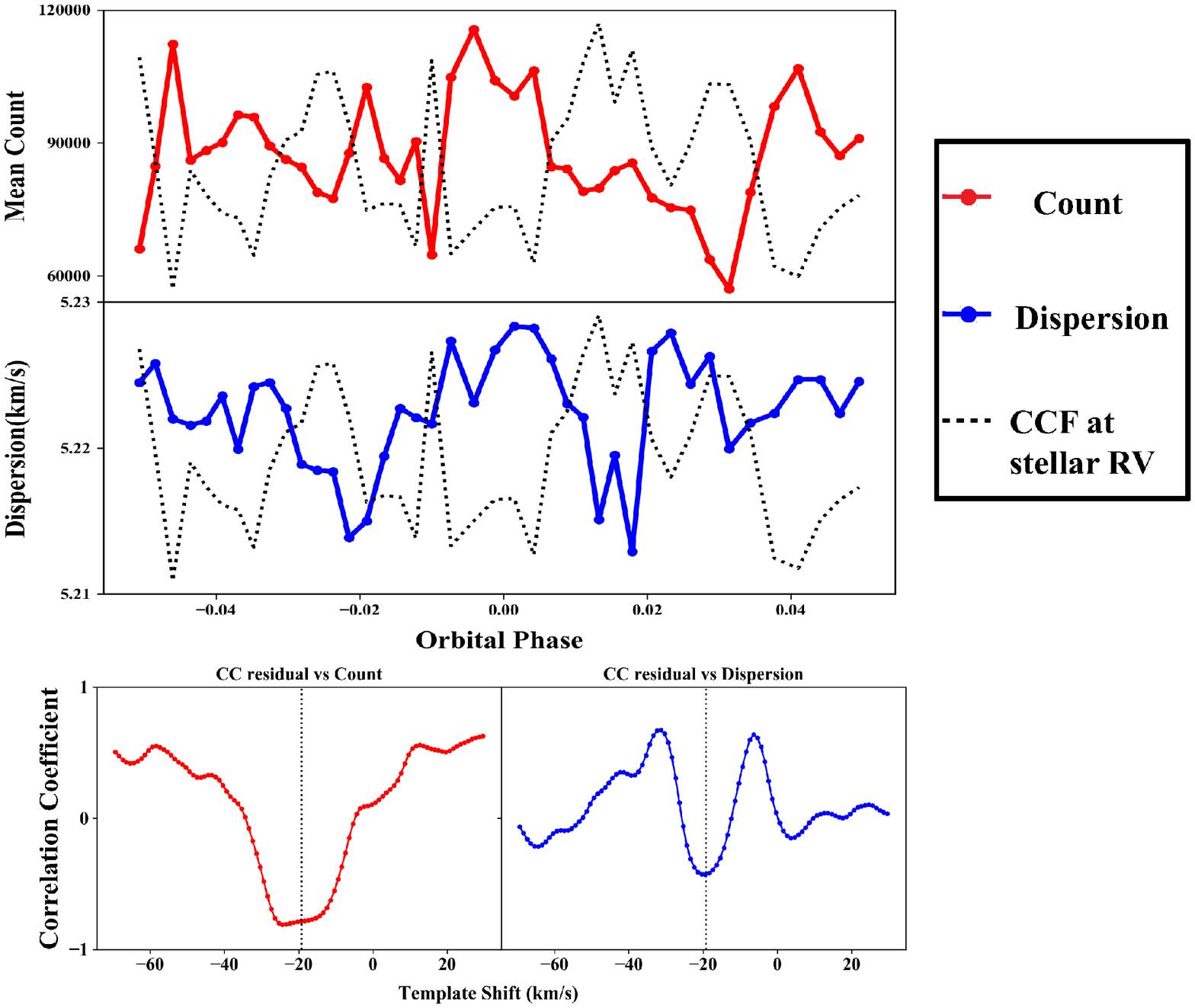}
\caption{Relationship between mean counts or dispersion in each frame and $C^{st}_\mathrm{norm}$.
The top and middle panels show the mean count (red) in 1 pixel (in 1-D spectrum) and dispersion (blue) in each frame. We fitted a Gaussian function to each row of $C^{st}_\mathrm{stellar}$ and used the width of the fitted Gaussian function as the dispersion in each frame. $C^{st}_\mathrm{norm}[i,\mathrm{RV_{sys}}]$, namely, the value of $C^{st}_\mathrm{norm}$ at the stellar radial velocity, is also plotted as a black-dotted line. The bottom figure shows the correlation coefficient for $C^{st}_\mathrm{norm}[i,\delta v]$ and the mean count (red) and dispersion (blue). $C^{st}_\mathrm{norm}$ at the stellar radial velocity were strongly correlated with the mean count, and  $C^{st}_\mathrm{norm}$ around the stellar radial velocity have correlations with the variations of the dispersion in each frame. \label{fig:noisecorr}}
\end{center}
\end{figure*}

We performed a correction of the stellar line profile and detrending of the quasi-static signals by the SYSREM algorithm as follows. In the transmission spectrum matrix, the dominant signal is the stellar lines. The line profile for the stellar spectrum is time-varying. We calculated a cross-correlation function matrix (CCF matrix) $C$ of the transmission spectrum matrix. $C$ is a function of the transmission spectrum matrix $F_\mathrm{xxx}$ and the stellar model spectrum $M$, and is defined as
\begin{equation}
C^{st}_\mathrm{xxx}[i,\delta v] = F_\mathrm{xxx}[i] \star M(\delta v),
\end{equation}
where $i$ denotes the frame number, $\star$ is the cross-correlation operator, and $M(\delta v)$ is the model spectrum shifted by $\delta v$.  We used a grid with steps of 1 ${\rm km \ s^{-1}}$ as $\delta v$. We denote $F_\mathrm{stellar}$ normalized by the mean of the out-transit as $F_\mathrm{norm}$. This quantity ideally becomes the transmission spectrum matrix without the stellar signals. However, we found significant residuals originating from stellar signals in $F_\mathrm{stellar}$. 
Figure \ref{fig:ccfresidual} shows the CCF matrices, $C^{st}_\mathrm{stellar}$ and $C^{st}_\mathrm{norm}$. Residuals remain in $C^{st}_\mathrm{norm}$ at the stellar radial velocity. The strong residuals in $C^{st}_\mathrm{norm}$ were correlated with the mean count and the resolution of each frame (Figure \ref{fig:noisecorr}). Therefore, we assume these residuals originate from an imperfect correction of the non-linearity of the CCD and instrumental profile (IP) variations. These residuals are problematic for detecting planetary signals because the stellar radial velocity is close to that of the planet in the planetary radial velocity--systemic velocity ($V_{sys}$)plane.

As a first step to correct these residuals, we performed a deconvolution of the IP in $S_\mathrm{raw}$. We estimated the IP of the observation night by a flat frame with an iodine cell, obtained just before the observation of the targets. 
We shifted each deconvolved spectrum and aligned them to the stellar rest frame, as shown in Section 2.4.
Then, we generated a template stellar spectrum by averaging the shifted spectra. We fitted 
\begin{eqnarray}
a \, [ T \ast {\rm IP} ](\lambda)
\end{eqnarray}
to each raw value of $F_\mathrm{stellar}$, where $a$ is a scaling factor, $T$ is the template stellar spectrum, and $\ast$ indicates the convolution operator. The scaling factor $a$ (for the non-linearity) and the IP are free parameters in this fitting. The IP was modeled by nine satellite Gaussian functions (\citealt{Valenti1995}). The positions and widths of the Gaussians were fixed. The only free parameters were their amplitudes. We normalized each spectrum by dividing each frame by the fitted template spectrum. The iodine cell generates numerous deep and sharp absorption lines in the 5000--5800 \AA \ region in the nineteen echelle orders (100 to 118) we used. We made this correction to these nineteen orders only. For the other orders, we normalized them by dividing by the mean spectrum. We denote the transmission spectrum matrix generated by this procedure by $F_\mathrm{IP}$. Figure \ref{fig:ccfresidual} also shows $C^{st}_\mathrm{IP}$, which is a CCF matrix generated from $F_\mathrm{IP}$ and the model stellar spectrum. As shown, the correction suppressed the noise at the stellar radial velocity.\\

Next, we used the SYSREM algorithm to remove the residuals from the stellar and telluric absorptions (\citealt{Tamuz2005}, \citealt{Mazeh2007}). SYSREM is an algorithm based on principal component subtraction and was developed to remove systematic variations in large numbers of transit lightcurve.
Previous studies have applied this technique to remove telluric and stellar absorption lines in high-resolution spectra and succeeded in detecting signals of planetary atmosphere (e.g., \citealt{Birkby2013,Birkky2017}, \citealt{Nugroho2017}, \citealt{Hawker2018}, \citealt{AlonsoF2019}, \citealt{Sanchez2019}). 
In our analysis, wavelength bins of the spectrum matrix were treated as numerous lightcurves.
The systematic variation was removed by minimizing 
\begin{equation}
S = \sum_{ij} \frac{(r_{ij}-c_i a_j)^2}{\sigma_{ij}^2}
\end{equation}
where $r_{ij}$ is the average-subtracted stellar magnitude of the wavelength bin $i$, $c_i$ is the effective extinction coefficient of the wavelength bin $i$, $a_j$ is the effective air mass of the $j$-th frame, and $\sigma_{ij}$ is the uncertainty of the wavelength bin $i$ of the $j$-th frame.
Each data point was weighted by its uncertainty, and the most significant systematic component was fitted and removed.
Therefore, $c_i$ and $a_j$ are not necessarily the real extinction coefficient and real air mass.
The SYSREM algorithm can remove not only the telluric and stellar absorption lines but also other systematic effects (e.g., variations of air mass or water vapor column density, variations in instrumental conditions) by iterating the SYSREM reduction to the residual. We applied SYSREM to $F_\mathrm{IP}$ with fifteen iterations.

Thus, we obtained 3-dimensional $F_\mathrm{SYS}[s,i,\lambda]$ for all of the echelle orders, with dimensions of sixteen (denoted by $s$ and ranging from 0 to 15 SYSREM iterations) $\times$ 41 (frame number, denoted by $i$)  $\times$ the number of wavelength bins. \\
The center position of the planetary absorption lines changes during observations according to the orbital motion, while the dominant noises from the stellar and telluric absorption lines are quasi-static. SYSREM removes this static noise, leaving the planetary signals. However, the planetary signals are also eliminated as the iterations proceed too much. When the S/N reaches the highest value, we regard the iteration number to be the optimized value. Because the noise level may be different for each order, the optimal SYSREM iteration number might be different for each order. The injection of artificial planetary signals was often employed (e.g., \citealt{Birkby2017}) to find  the optimal iteration number for each order, although this method may result in some bias (\citealt{Cabot2019}). Thus, recent studies use common iteration number for all orders with highest S/N (e.g., \citealt{Turner2020}, \citealt{Nugroho2020}). Therefore, we determined a common iteration number for all the orders with the highest S/N. 

\section{Cross-Correlation Analysis}
\subsection{Model spectra for the cross-correlation analysis}
We generated the model transmission spectra for the cross-correlation template of Sc I, Ti I, V I, Cr I, Mn I, Fe I, and Co I for atomic species, and TiO for molecular species. These atoms/molecules have many absorption lines in the wavelength range of our data. Each model spectrum included absorption by a single atom or molecule plus continuum absorption. We included a continuum absorption cross section of ${\rm H^{-}}$ and Rayleigh scattering by ${\rm H \ and \  H_{2}}$.  Here, we describe the procedure to generate them. \\
We assumed a cloud-free, isothermal atmosphere at a temperature of 2000 K. We calculated the cross section using {\it Python for Computational ATmospheric Spectroscopy} (Py4CAtS, \citealt{Schreier2019}) for atomic species. 
We used partition functions taken from \citet{Barklem_Collet2016} with a spline interpolation to obtain the partition function at the atmospheric temperature. 
We used the line list given by \citet{Kurucz2018}. 
For TiO, we used HELIOS-K (\citealt{Grimm_Heng2015}) to compute the cross section. We used two different line lists for TiO, provided by \citet{Plez1998} (updated and corrected in 2012, hereafter TiO-Plez) and \citet{TiOtoto} (hereafter TiO-Toto). We considered the five most abundant isotopes, ${\rm ^{46}Ti^{16}O, \ ^{47}Ti^{16}O, \ ^{48}Ti^{16}O, \ ^{49}Ti^{16}O, \ ^{50}Ti^{16}O }$ and combined them with the natural isotope ratio.

The abundance of each atom and molecule in each layer was calculated using {\it FastChem} (\citealt{Stock2018}), which calculates the abundances of atoms and molecules in the chemical equilibrium. We assumed a metallicity of 0.36 (\citealt{Sato2005}). 
We computed the line profile based on the Voigt profile with natural broadening and thermal broadening, because the transmission spectroscopy mainly probes a low-pressure region of the planetary atmosphere and thus the contribution of pressure-broadening is much smaller than that for other broadening mechanisms. We assumed a 1-D plane-parallel atmosphere with one hundred layers, which were evenly spaced in the log pressure scale from 10 to ${\rm 10^{-15}}$ bar.\\
The planetary radius at a wavelength of $\lambda$, $R_p (\lambda)$, and the model transmission spectrum $ Tr(\lambda)$ were calculated by

\begin{eqnarray}
R_p^2 (\lambda) &=& R_{p_{0}}^2+dR_p^2 (\lambda) \nonumber \\
&=& R_{p_{0}}^2 + 2\int_{R_{p_{0}}}^{R_{p_{0}}+R_{{\rm max}}} (1-\exp(-\tilde{\tau}_{\lambda,\tilde{r}}))\tilde{r} d\tilde{r} \label{eq:drp}\\
Tr(\lambda) &=& {\rm 1 - }\Bigl( \frac{R_{p}(\lambda)}{R_{s}} \Bigr)^{2}
\end{eqnarray}
where $R_{p_{0}}$ is the planetary radius with white light, $R_{{\rm max}}$ is the maximum height of the model atmosphere, $\tau_{\lambda,r}$ is the optical depth at $\lambda$ and at the coordinates $r$, and $R_s$ is the stellar radius. The symbols with tildes indicate physical quantities integrated over the transmission chord direction. 

\subsection{Cross-correlation and SN map}

We generated model spectra with a radial velocity shift from $-$262.5 ${\rm km \ s^{-1}}$ to 224.5 ${\rm km \ s^{-1}}$ with a step size of 1 ${\rm km \ s^{-1}}$, corresponding to a systemic velocity ($V_{sys}$) ranging from $-$120 ${\rm km \ s^{-1}}$ to 80 ${\rm km \ s^{-1}}$ and the semi-amplitude of the planetary radial velocity ($K_{p}$) from $-$100 to 300 ${\rm km \ s^{-1}}$ (see equation \ref{eq:rv}). 

We computed $C^\mathrm{atom}[s,i,\delta v]$ as a function of $F_\mathrm{SYS}$ and a shifted model spectrum of each atom (denoted by $M_\mathrm{atom}$). 
We have subtracted the continuum of model spectra to zero. We applied two high-pass filters to the spectrum of each frame to remove the effects of low-order variations before computing the cross-correlation. The cross-correlation was computed by
\begin{equation}
C^\mathrm{atom}[s,i,\delta v] = \sum_{k} \frac{F_\mathrm{SYS}[s,i,k] M_\mathrm{atom}(\delta v)[k]}{\sigma_k^2},
\end{equation}
where $\sigma_k$ is the error value at the $k$-th wavelength bin. We assumed that $\sigma_k$ is the square root of the sum of the variance of each wavelength bin and frame of $F_\mathrm{norm}$.
Then, we summed $C^\mathrm{atom}$ for the orders in which the planetary signal is strong. 
The criterion for this selection process is that the strongest line depth in the order is deeper than ${\rm 2e^{-5}}$ to the continuum, which corresponds to $dR_p \sim \frac{H}{2}$ in equation \ref{eq:drp}, where $H$ is the scale height of the planetary atmosphere. . 
In this way, we obtained $C^\mathrm{atom}$, whose dimensions are 41 (frame number) $\times$ 488 (radial velocity from $-$262.5 ${\rm km \ s^{-1}}$ to 224.5 ${\rm km \ s^{-1}}$) for each SYSREM iteration number. We computed the CCF map in the $K_{p}$-$V_{sys}$ plane as follows: we integrated $C^\mathrm{atom}[s]$ (in the frame--radial velocity plane) along the path of expected planetary radial velocity $RV_{p}$,
\begin{equation}
RV_{p} = K_{p}\ \sin (2 \pi \phi(t)) +V_{sys}
\label{eq:rv}
\end{equation}
\begin{equation}
\phi (t) = \frac{t-T_{o}}{P},
\end{equation}
where $\phi (t)$ is the orbital phase of the planet, $t$ is the Barycentric Julian Day (BJD) of each exposure. We only used the in-transit frames for this calculation. The orbital phase of each frame was calculated by the system parameters (Table 2) using {\sl batman} python software (\citealt{Laura2015}). 
We also calculated the expected amount of starlight blocked by the planet for each frame by {\sl batman}, and used it as a weighting coefficient with a $K_{p}$--$V_{sys}$ integration. To make the S/N map, we divide the CCF map by the standard deviation of the map after excluding the expected position of the planetary signal as well as the region with $|K_p|$ $<$ 50 ${\rm km \ s^{-1}}$, since the signal with low $|K_p|$ approaches zero as SYSREM iteration proceeds.

To estimate the errors in $K_{p}$ and $V_{sys}$ for the detected peaks, we calculated the likelihood maps (\citealt{Brogi_Line2019}, \citealt{Gibson2020}, \citealt{Nugroho2020}) in the $K_{p}$--$V_{sys}$ plane. The log likelihood  is defined as
\begin{equation}
\log{L} =  -\frac{N}{2}\ln{\left[ \frac{1}{N} \sum_{i}\left(\frac{f_i^2}{\sigma_i^2}+\frac{m_i^2}{\sigma_i^2}-2\frac{f_i m_i}{\sigma_i^2}\right) \right]},
\end{equation}
where $N$ is the number of data points. The third term can be represented by the CCF maps in the $K_{p}$--$V_{sys}$ plane. The likelihood map is calculated with the exponential of $\log{L}$ after subtracting its maximum value. We fitted a Gaussian function to the slice of the likelihood map at the maximum position. Finally, the errors in $K_{p}$ and $V_{sys}$ were estimated from the standard deviation of the best-fit Gaussian function. We also calculated the detection significance by the phase shuffling method (e.g., \citealt{Esteves2017}). We randomly shuffled the in-transit CCF map in the phase axis and integrated them with $K_{p}$ of the detected peak. We repeated this calculation 10000 times. The noise at each $V_{sys}$ was estimated by the standard deviation of the integrated CCF value.

\section{Results}
\begin{figure*}[th]
\begin{center}
\includegraphics[width=18cm]{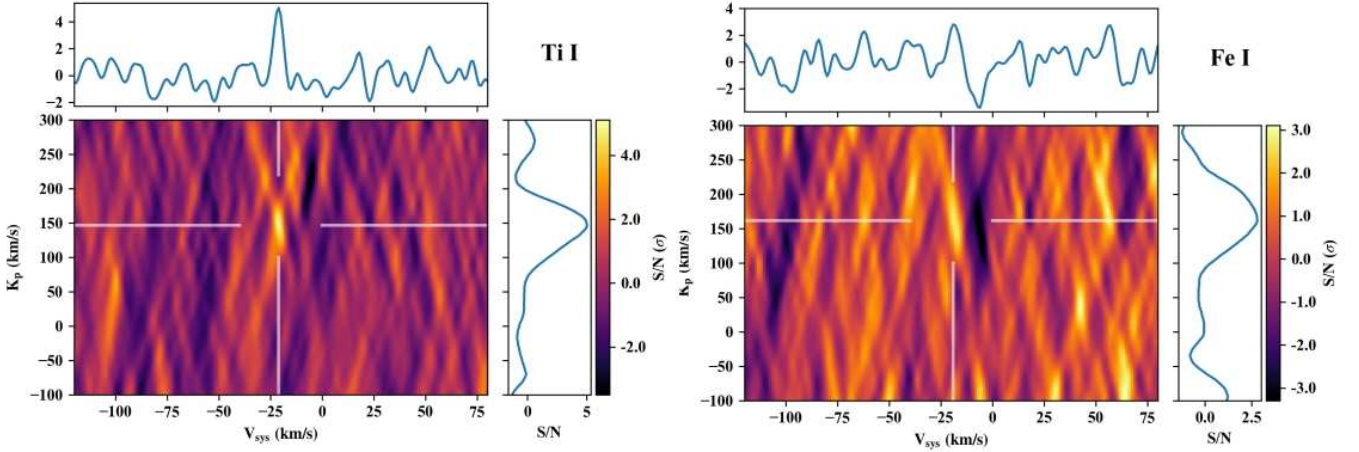}
\caption{S/N map of Ti I and Fe I in $K_{p}$--$V_{sys}$ plane. 1-D cross sections along the $K_{p}$ and $V_{sys}$ directions are also shown.}
\label{fig:result1}
\end{center}
\end{figure*}

\begin{figure}
\begin{center}
\includegraphics[width=9cm]{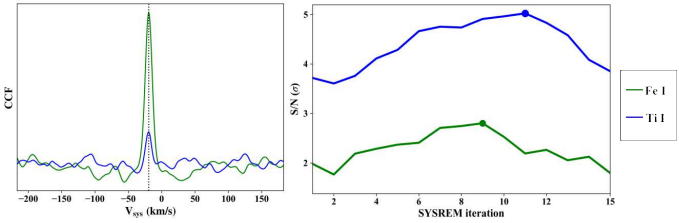}
\caption{(left) Cross-correlation function of model transmission template spectrum and $F_\mathrm{stellar}$. The black dotted line shows the stellar radial velocity. There are  also absorption lines for Fe I and Ti I in the stellar spectrum; the absorption by Fe I is stronger than that of Ti I.
(right) S/N for detected peak with SYSREM iteration number. The adopted iteration numbers for the S/N map in Figure 5 (the best iteration number) are shown by dots. }
\label{fig:sysite}
\end{center}
\end{figure}

\begin{figure*} 
\begin{center}
\includegraphics[width=18cm]{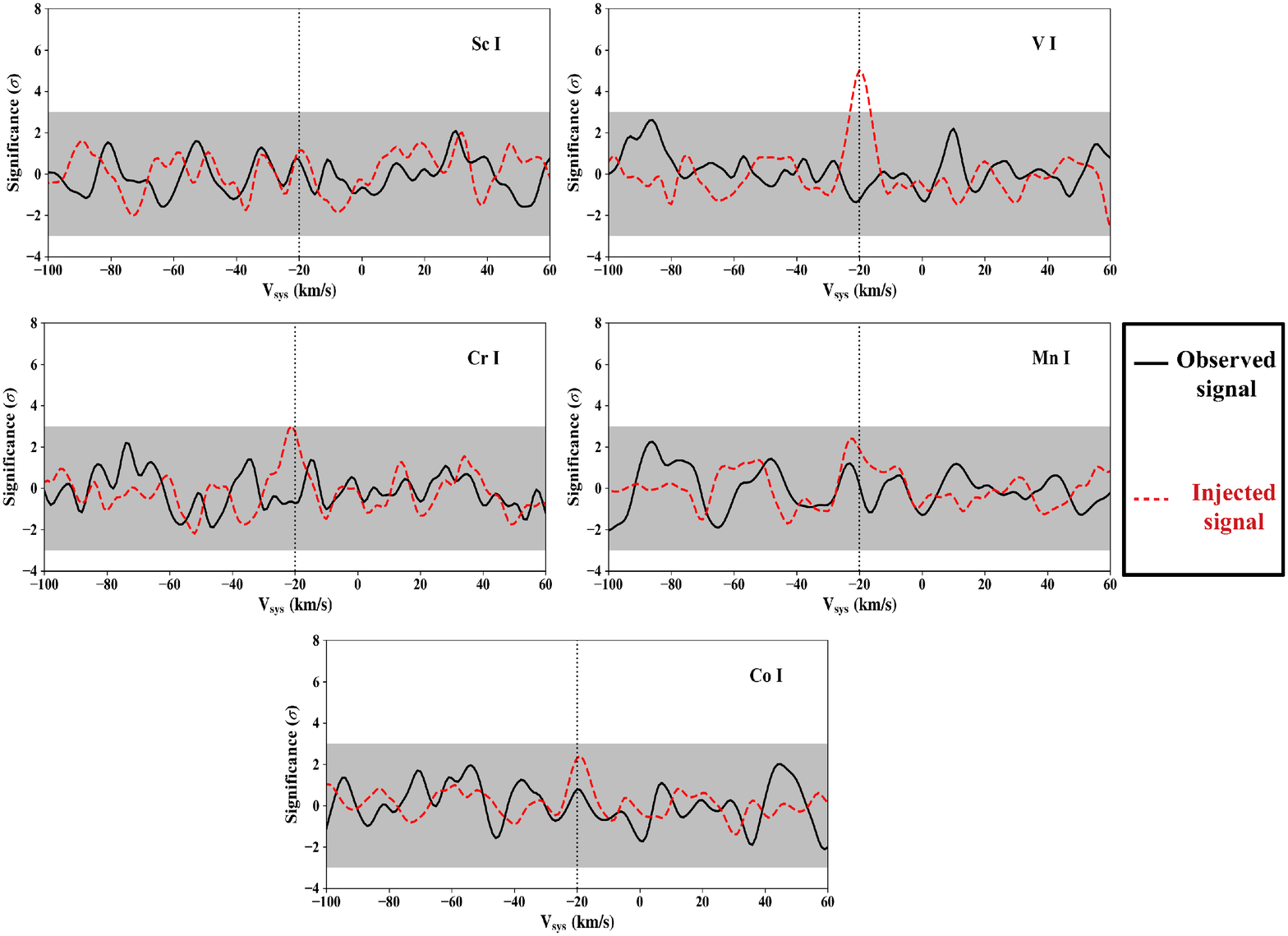}
\caption{Observed signals and results of injection tests for non-detected atomic and molecular species except TiO. The red dotted line shows the significance of the injected signals at $K_{p} = $-150 ${\rm km \ s^{-1}}$ and $V_{sys}$ = $-$20 ${\rm km \ s^{-1}}$ after 10 SYSREM iterations. The black solid line shows the significance of the observed signals at $K_{p} = $ 150 ${\rm km \ s^{-1}}$ after 10 SYSREM iterations. The grey shading represents the $\sigma$ $<$ 3 detection limit.}
\label{fig:result2}
\end{center}
\end{figure*}

\subsection{Neutral iron and titanium}

We made the first detection of neutral titanium (Ti I) in a planetary atmosphere. 
We also detected a marginal signal of neutral iron (Fe I). The S/N for the detected signals was 5.0 for Ti I and 2.8 for Fe I. The detection significances calculated by the phase shuffling method were 4.4 and 2.8.

Figure \ref{fig:result1} shows the S/N map in the $K_{p}$--$V_{sys}$ plane. The noise level was estimated from the standard deviation of the CCF values excluding the region around the expected positions of the planetary signal. 

The peak values are $K_{p} {\rm = 146.9 \pm ^{5.1}_{5.1}}$ ${\rm km \ s^{-1}}$, $\Delta V{\rm = -3.2 \pm ^{0.4}_{0.4}}$ ${\rm km \ s^{-1}}$ for Ti I, and $K_{p} {\rm = 163.4 \pm ^{7.8}_{7.8}}$ ${\rm km \ s^{-1}}$,$\Delta V{\rm = -0.7 \pm ^{0.6}_{0.6}}$ ${\rm km \ s^{-1}}$ for Fe I, respectively. The values of $\Delta V$ means the difference between $V_{sys}$ of the detected peak and the stellar radial velocity (-17.92 ${\rm km \ s^{-1}}$). 
We note that we only used echelle orders 100 to 118 for Fe I, because the stellar noise was significant in orders without the line profile correction (see Section 2.5). 
We excluded the frames 22 and 23 (the two closest frames to the mid-transit) for both atomic species in order to avoid affects by the residual of the stellar signal, because the planetary signal was expected to be overlapped with the center of the stellar absorption lines in these frames. 
Figure \ref{fig:sysite} shows the S/N for the detected peak with SYSREM iteration numbers from 0 to 15. 
The SYSREM iteration number for the highest S/N is 9 for Fe I and 11 for Ti I.
Figure \ref{fig:sysite} also shows the cross-correlation functions for the model transmission template spectrum and $F_\mathrm{stellar}$. 
Figure \ref{fig:sysite} shows that the absorption lines for Fe I in the stellar spectrum were stronger than those for Ti I. 
The marginal detection of Fe I might be due to the strong contamination from stellar spectrum.

\begin{table}
  \centering
  \begin{tabular}{c|c|c|c|c}
    species & $K_{p}$& $\Delta V$& S/N & $\sigma$ \\ \hline\hline
    Fe I &${\rm = 163.4 \pm ^{7.8}_{7.8}}$&${\rm = -0.7 \pm ^{0.6}_{0.6}}$&2.8 & 2.8 \\
    Ti I &$ {\rm = 146.9\pm ^{5.1}_{5.1}}$& ${\rm = -3.2 \pm ^{0.4}_{0.4}}$ & 5.0 & 4.4
  \end{tabular}
  \caption{Positions, S/N, and significance of detected peaks in the $K_{p}$--$V_{sys}$ plane. $\Delta V$ is the difference between $V_{sys}$ for the detected peak and the stellar radial velocity ($-$17.92 ${\rm km \ s^{-1}}$). The detection significance was estimated by the phase-shuffling method (\citealt{Esteves2017}).}
  \label{tb:result}
\end{table}

\begin{figure}
\begin{center}
\includegraphics[width=9cm]{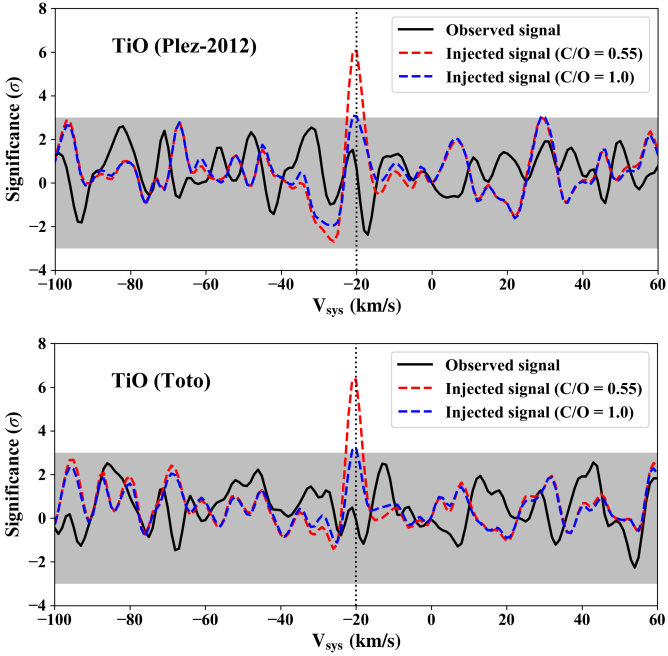}
\caption{Result of injection test of TiO under chemical equilibrium with a C/O value of 0.55 (solar) and 1.0 after 10 SYSREM iterations. The top figure is the result with the linelist provided by \citet{Plez1998}, and the bottom figure is the result with the linelist from \citet{TiOtoto}. We can detect TiO with $\sigma$ $>$ 6, if the abundance of TiO is the same as the calculated value with the solar C/O ratio. However, if the C/O ratio is enhanced to 1, the planetary signal is expected to be at the noise level.}
\label{fig:tio}
\end{center}
\end{figure}

\begin{figure}
\begin{center}
\includegraphics[width=9cm]{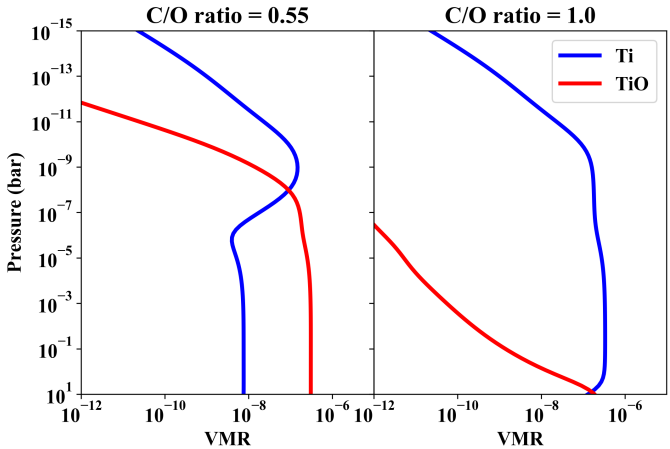}
\caption{Volume mixing ratio of Ti I and TiO with a temperature of 2000 K calculated by {\it Fastchem}. We set the metallicity to 0.36.}
\label{fig:titio}
\end{center}
\end{figure}

\begin{figure}
\begin{center}
\includegraphics[width=9cm]{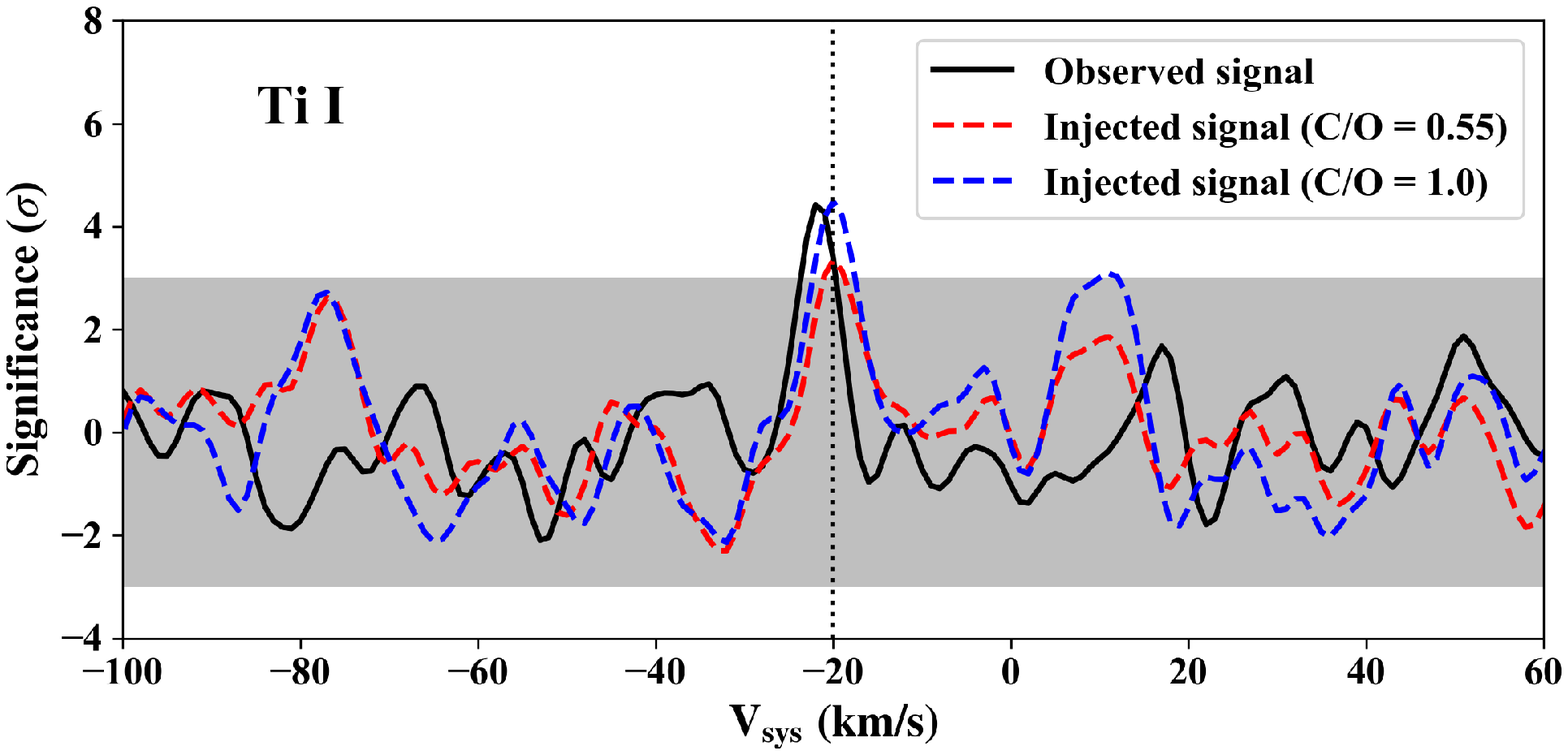}
\caption{Observed signal and the result of an injection test of Ti I under chemical equilibrium with a C/O value of 0.55 (solar) and 1.0 after 11 SYSREM iterations. The significance of the observed Ti I signal is 4.4. }
\label{fig:injTi}
\end{center}
\end{figure}

\subsection{Non-detection of the other atomic/molecular species}
We could not find planetary signals for any other atomic or molecular species. For the non-detected species, we conducted an injection test to investigate the detection limit in our analysis (Figure \ref{fig:result2}, \ref{fig:tio}). We injected a model template spectrum of each of the species to the spectrum matrix before SYSREM detrending. We convoluted a model spectrum with a rotation kernel to simulate line broadening by planetary tidally locked rotation by $\sim$ 1.3 ${\rm km \ s^{-1}}$, and then injected the broadened model spectrum assuming $K_{p}$ = $-$150 ${\rm km \ s^{-1}}$ and  $V_{sys}$ = $-$20 ${\rm km \ s^{-1}}$. To avoid the effect of unseen planetary signals from the planetary atmosphere, $K_{p}$ was inverted (\citealt{Merritt2020}). The detection significance for the injected signal was calculated by the phase shuffling method. In this paper, we consider species to be undetectable with our analysis if the significance of the injected signal is $\lesssim$ 3 $\sigma$. 
For Sc I, Cr I, Mn I, and Co I, the expected signal was $\lesssim$ 3 $\sigma$, comparable to the noise, and thus we could not confirm the existence of these atomic/molecular species. 
Our analysis was able to detect V I with $\gtrsim$ 5 $\sigma$, showing that V I is depleted in the HD149026b atmosphere with $\gtrsim$ 5 $\sigma$, with the assumption of a cloud-free, 2000-K isothermal atmosphere, and chemical equilibrium. \\

\subsubsection{Titanium oxide}
Assuming a solar-like abundance and chemical equilibrium, titanium mainly exists in the form of titanium oxide (TiO) at a temperature of ${\rm \sim}$ 2000 K. Figure \ref{fig:titio} shows the abundance of Ti I and TiO in chemical equilibrium calculated using {\it Fastchem}.  We note the TiO-Toto linelist was confirmed to be accurate by cross-correlation with M dwarf's spectrum for $\lambda > $ 4400 \AA (\citealt{TiOtoto}). To consider the implication of the non-detection of TiO in our analysis for both linelists, TiO-Plez and TiO-Toto, we performed injection tests for TiO.

As shown in Figure \ref{fig:tio}, if the planet had a cloud-free, 2000-K isothermal atmosphere, and chemical equilibrium, our analysis could detect TiO signals with $>$ 6 $\sigma$ for both line lists assuming the line lists are completely accurate. This is indicative because we detected neutral Ti but not TiO. We discuss the implications in the next section in more detail.

\section{Discussion and Summary}

\subsection{Physical properties of the atmosphere of HD149026b}

Both Fe and Ti can condense and take the form of clouds below the condensation temperature. The condensation temperatures for typical condensates at a pressure of 1 mbar are 1650 K for FeO and 1582 K for $\mathrm{CaTiO_3}$ (\citealt{WakeSing2015}).  Our detection of Ti I (and the hint of Fe I) in a gaseous phase indicates that the temperature of HD149026b is higher than the condensation temperatures for Ti-bearing and Fe-bearing condensates. We note that the equilibrium temperature for HD149026b is $\sim$ 1680 K, consistent with our results, if we assume the received energy is evenly redistributed to the whole planet and the Bond albedo = 0.1.
This temperature is higher than the condensation temperature for most chemical species which can form clouds (\citealt{WakeSing2015}). 
Given this high temperature and the fact that we detected atomic species in the atmosphere, the effect of clouds would not be significant for HD 149026b. However, we cannot rule out the cloud hypothesis because several other chemical species have condensation temperatures higher than $\geq$ 2000 K (e.g., see Figure 2 in \citealt{Mbarek2016}).

A possible explanation for the non-detection of TiO is a supersolar C/O ratio in the planetary atmosphere, as reported for several other hot Jupiters (\citealt{Brewer2017}), as the abundance of TiO strongly depends on the C/O ratio. 
Other scenarios which have been proposed to explain the absence of TiO in the planetary atmospheres are less likely in this case as described below.
Our detection shows that titanium is not depleted in the atmosphere of HD149026b, thus titanium depletion by cold-trap effect (\citealt{Spiegel2009}, \citealt{Parmentier2013}) would not be significant in the atmosphere of HD149026b.
Strong chromospheric activity of the host star might destroy the TiO in planetary atmospheres, though the activity of HD149026 is low based on the analysis of Ca I\hspace{-.1em}I H \& K lines (\citealt{Knutson2010}).
Figure \ref{fig:titio} also shows the VMR of Ti I and TiO with C/O values of 0.55 and 1.0.
As the C/O ratio increases, the abundance of TiO significantly decreases. 
Figure \ref{fig:tio} shows the result of an injection test assuming a C/O ratio of 0.55 and 1.0. 
In the case of a C/O of 1.0, our analysis will not detect a TiO signal with $\lesssim$ 3 $\sigma$, therefore the scenario of C/O $\gtrsim$ 1 is consistent with our results. 
We also estimated the expected significance of the Ti I signal in the same way (Figure \ref{fig:injTi}). 
The predicted significance of Ti I with a C/O value of 0.55 and 1.0 was $\sim$ 3.3 and 4.5, respectively.
The significance of the detected signal was 4.4, thus our result is consistent with the scenario that the C/O ratio is higher than the solar value.
However, we note that the predicted significance strongly depends on the atmospheric properties such as thermal profile or scattering properties such as clouds, because the depth of absorption lines in the model transmission spectrum is affected by these properties. \\

The C/O ratio for the planetary atmosphere provides information on its formation history. Some hot Jupiters have been reported to have supersolar C/O ratios (e.g., HD209458b, \citealt{Mad_Sea2009}). A supersolar C/O ratio suggests that the planet swept up a large portion of its atmosphere from the gas in the protoplanetary disk outside the water snowline (gaseous oxygen is reduced); that is, the planet formed outside the water snowline and migrated to the current orbit (e.g., \citealt{Oberg2011}). Therefore, our detection of Ti I and non-detection of TiO suggest that HD149026b formed outside the disk and migrated to the current orbit. HD149026b is considered to have a massive core ($\sim$ 67 $\mathrm{M_\oplus}$) like WASP-59b (\citealt{Hebrard2013}). 
The formation history of hot Jupiters with massive cores is still in debate. It is considered that such hot Jupiters are formed in an environment with a high dust grain density (\citealt{Ikoma2006}). 
\citet{Kanagawa2018} showed that a planetary-induced gap in the protoplanetary disk can make a 'dust ring' at the outer edge of the disk, and a gas giant with a massive core can be formed in the dust ring. 
In this scenario, HD149026b would have migrated by dynamical scattering. 
Our detection of neutral metals and the implication of a high C/O ratio provide clues on the formation history of hot Jupiters with massive cores. \\
\citet{Line2014} reported that the atmospheric C/O ratio for HD149026b is 0.45 to 1.0 (68\% confidence), determined by {\sl Spitzer} photometry of a secondary eclipse. 
The C/O ratio of $\sim $ 1 hypothesis does not contradict their results. 
However, the detection limit for atom/molecule abundance discussed in this paper depends on the properties of the planetary atmosphere, such as scattering and vertical mixing. 
In our model we assumed Rayleigh scattering by ${\rm H \ and \  H_{2}}$, continuum absorption by ${\rm H^-}$, and a cloud-free, isothermal atmosphere under chemical equilibrium, and the detection limits we set should be interpreted in this context only. \\

\subsection{Origin of the blueshift}
The detected signals were slightly blue-shifted compared with the central star. 
A typical wind speed from the dayside to nightside in a tidally locked hot Jupiter's atmosphere is a few ${\rm km \ s^{-1}}$ (e.g., \citealt{Kataria2016}), which is consistent with the blueshift of the detected signals. However, if we assume the planet is tidally locked, the rotational velocity at the equator is $\sim$ 1.3 ${\rm km \ s^{-1}}$. The asymmetric distribution of atomic species in the planetary atmosphere also shifts the planetary signals (\citealt{Ehrenreich2020}). Since the blueshift of the detected signals is also consistent with the planetary rotation, we could not confirm the cause of the blueshift of the detected signals.

\subsection{Summary}
In this paper, we presented the results of high-resolution transmission spectroscopy of HD149026b at visible wavelengths using Subaru/HDS. We detected the signal of neutral titanium (4.4 $\sigma$) and marginal signal of neutral iron (2.8 $\sigma$) in the atmosphere of HD149026b. Our study is the first detection of neutral titanium in the planetary atmosphere. 

The detection of Ti I and the non-detection of TiO suggests a supersolar C/O ratio in the atmosphere of HD149026b.\\
The formation scenario for hot Jupiters is of great interest in exoplanetary science. HD149026b is a hot Jupiter with a massive core, and its formation history is still controversial. Our detection of neutral metals and the implication of a supersolar C/O ratio add important clues for understanding the unique properties of this hot Jupiter.
\\

 We wish to thank Takayuki Kotani for meaningful advice. We are also grateful to Akito Tajitsu for providing valuable information on the HDS reduction. This work was supported by JSPS KAKENHI grant numbers 19J11805 (M.I.), JP18H04577, JP18H01247, JP20H00170 (H.K.), and JP18H05442 (MT). This work was also supported by the JSPS Core-to-Core Program Planet2 and SATELLITE Research from the Astrobiology Center (AB022006). S.K.N. would like to acknowledge support from the UK Science Technology and Facility Council grant ST/P000312/1. 
 Y.K. is supported by the European Union's Horizon 2020 Research and Innovation Programme under Grant Agreement 776403.
 The authors wish to recognize
and acknowledge the very significant cultural role and
reverence that the summit of Mauna Kea has always
had within the indigenous Hawaiian community. We
are most fortunate to have the opportunity to conduct
observations from this mountain.
\bibliographystyle{aasjournal}
\bibliography{sample63}

\begin{thebibliography}{}
\expandafter\ifx\csname natexlab\endcsname\relax\def\natexlab#1{#1}\fi
\providecommand{\url}[1]{\href{#1}{#1}}
\providecommand{\dodoi}[1]{doi:~\href{http://doi.org/#1}{\nolinkurl{#1}}}
\providecommand{\doeprint}[1]{\href{http://ascl.net/#1}{\nolinkurl{http://ascl.net/#1}}}
\providecommand{\doarXiv}[1]{\href{https://arxiv.org/abs/#1}{\nolinkurl{https://arxiv.org/abs/#1}}}

\bibitem[{{Alonso-Floriano} {et~al.}(2019){Alonso-Floriano},
  {S{\'a}nchez-L{\'o}pez}, {Snellen}, {L{\'o}pez-Puertas}, {Nagel}, {Amado},
  {Bauer}, {Caballero}, {Czesla}, {Nortmann}, {Pall{\'e}}, {Salz}, {Reiners},
  {Ribas}, {Quirrenbach}, {Aceituno}, {Anglada-Escud{\'e}}, {B{\'e}jar},
  {Guenther}, {Henning}, {Kaminski}, {K{\"u}rster}, {Lamp{\'o}n}, {Lara},
  {Montes}, {Morales}, {Tal-Or}, {Schmitt}, {Zapatero Osorio}, \&
  {Zechmeister}}]{AlonsoF2019}
{Alonso-Floriano}, F.~J., {S{\'a}nchez-L{\'o}pez}, A., {Snellen}, I.~A.~G.,
  {et~al.} 2019, \aap, 621, A74, \dodoi{10.1051/0004-6361/201834339}

\bibitem[{{Baba} {et~al.}(2002){Baba}, {Yasuda}, {Ichikawa}, {Yagi}, {Iwamoto},
  {Takata}, {Horaguchi}, {Taga}, {Watanabe}, {Ozawa}, \& {Hamabe}}]{Baba2002}
{Baba}, H., {Yasuda}, N., {Ichikawa}, S.-I., {et~al.} 2002, Astronomical
  Society of the Pacific Conference Series, Vol. 281, {Development of the
  Subaru-Mitaka-Okayama-Kiso Archive System}, ed. D.~A. {Bohlender},
  D.~{Durand}, \& T.~H. {Handley}, 298

\bibitem[{{Barklem} \& {Collet}(2016)}]{Barklem_Collet2016}
{Barklem}, P.~S., \& {Collet}, R. 2016, \aap, 588, A96,
  \dodoi{10.1051/0004-6361/201526961}

\bibitem[{{Batygin} {et~al.}(2016){Batygin}, {Bodenheimer}, \&
  {Laughlin}}]{Batygin2016}
{Batygin}, K., {Bodenheimer}, P.~H., \& {Laughlin}, G.~P. 2016, \apj, 829, 114,
  \dodoi{10.3847/0004-637X/829/2/114}

\bibitem[{{Birkby} {et~al.}(2013){Birkby}, {de Kok}, {Brogi}, {de Mooij},
  {Schwarz}, {Albrecht}, \& {Snellen}}]{Birkby2013}
{Birkby}, J.~L., {de Kok}, R.~J., {Brogi}, M., {et~al.} 2013, \mnras, 436, L35,
  \dodoi{10.1093/mnrasl/slt107}

\bibitem[{{Birkby} {et~al.}(2017{\natexlab{a}}){Birkby}, {de Kok}, {Brogi},
  {Schwarz}, \& {Snellen}}]{Birkky2017}
{Birkby}, J.~L., {de Kok}, R.~J., {Brogi}, M., {Schwarz}, H., \& {Snellen},
  I.~A.~G. 2017{\natexlab{a}}, \aj, 153, 138, \dodoi{10.3847/1538-3881/aa5c87}

\bibitem[{{Birkby} {et~al.}(2017{\natexlab{b}}){Birkby}, {de Kok}, {Brogi},
  {Schwarz}, \& {Snellen}}]{Birkby2017}
---. 2017{\natexlab{b}}, \aj, 153, 138, \dodoi{10.3847/1538-3881/aa5c87}

\bibitem[{{Boley} {et~al.}(2016){Boley}, {Granados Contreras}, \&
  {Gladman}}]{Boley2016A}
{Boley}, A.~C., {Granados Contreras}, A.~P., \& {Gladman}, B. 2016, \apjl, 817,
  L17, \dodoi{10.3847/2041-8205/817/2/L17}

\bibitem[{{Bonomo} {et~al.}(2017){Bonomo}, {Desidera}, {Benatti}, {Borsa},
  {Crespi}, {Damasso}, {Lanza}, {Sozzetti}, {Lodato}, {Marzari}, {Boccato},
  {Claudi}, {Cosentino}, {Covino}, {Gratton}, {Maggio}, {Micela}, {Molinari},
  {Pagano}, {Piotto}, {Poretti}, {Smareglia}, {Affer}, {Biazzo}, {Bignamini},
  {Esposito}, {Giacobbe}, {H{\'e}brard}, {Malavolta}, {Maldonado}, {Mancini},
  {Martinez Fiorenzano}, {Masiero}, {Nascimbeni}, {Pedani}, {Rainer}, \& {Scand
  ariato}}]{Bonomo2017}
{Bonomo}, A.~S., {Desidera}, S., {Benatti}, S., {et~al.} 2017, \aap, 602, A107,
  \dodoi{10.1051/0004-6361/201629882}

\bibitem[{{Brewer} {et~al.}(2017){Brewer}, {Fischer}, \&
  {Madhusudhan}}]{Brewer2017}
{Brewer}, J.~M., {Fischer}, D.~A., \& {Madhusudhan}, N. 2017, \aj, 153, 83,
  \dodoi{10.3847/1538-3881/153/2/83}

\bibitem[{{Brogi} \& {Line}(2019)}]{Brogi_Line2019}
{Brogi}, M., \& {Line}, M.~R. 2019, \aj, 157, 114,
  \dodoi{10.3847/1538-3881/aaffd3}

\bibitem[{{Brogi} {et~al.}(2012){Brogi}, {Snellen}, {de Kok}, {Albrecht},
  {Birkby}, \& {de Mooij}}]{Brogi2012}
{Brogi}, M., {Snellen}, I. A.~G., {de Kok}, R.~J., {et~al.} 2012, \nat, 486,
  502, \dodoi{10.1038/nature11161}

\bibitem[{{Cabot} {et~al.}(2019){Cabot}, {Madhusudhan}, {Hawker}, \&
  {Gandhi}}]{Cabot2019}
{Cabot}, S. H.~C., {Madhusudhan}, N., {Hawker}, G.~A., \& {Gandhi}, S. 2019,
  \mnras, 482, 4422, \dodoi{10.1093/mnras/sty2994}

\bibitem[{{Casasayas-Barris} {et~al.}(2018){Casasayas-Barris}, {Pall{\'e}},
  {Yan}, {Chen}, {Albrecht}, {Nortmann}, {Van Eylen}, {Snellen}, {Talens},
  {Gonz{\'a}lez Hern{\'a}ndez}, {Rebolo}, \& {Otten}}]{Casasayas2018}
{Casasayas-Barris}, N., {Pall{\'e}}, E., {Yan}, F., {et~al.} 2018, \aap, 616,
  A151, \dodoi{10.1051/0004-6361/201832963}

\bibitem[{{Casasayas-Barris} {et~al.}(2019){Casasayas-Barris}, {Pall{\'e}},
  {Yan}, {Chen}, {Kohl}, {Stangret}, {Parviainen}, {Helling}, {Watanabe},
  {Czesla}, {Fukui}, {Monta{\~n}{\'e}s-Rodr{\'\i}guez}, {Nagel}, {Narita},
  {Nortmann}, {Nowak}, {Schmitt}, \& {Zapatero Osorio}}]{Casasayas2019}
---. 2019, \aap, 628, A9, \dodoi{10.1051/0004-6361/201935623}

\bibitem[{{Charbonneau} {et~al.}(2008){Charbonneau}, {Knutson}, {Barman},
  {Allen}, {Mayor}, {Megeath}, {Queloz}, \& {Udry}}]{Charbonneau2008}
{Charbonneau}, D., {Knutson}, H.~A., {Barman}, T., {et~al.} 2008, \apj, 686,
  1341, \dodoi{10.1086/591635}

\bibitem[{{Coelho} {et~al.}(2005){Coelho}, {Barbuy}, {Mel{\'e}ndez},
  {Schiavon}, \& {Castilho}}]{Coelho2005}
{Coelho}, P., {Barbuy}, B., {Mel{\'e}ndez}, J., {Schiavon}, R.~P., \&
  {Castilho}, B.~V. 2005, \aap, 443, 735, \dodoi{10.1051/0004-6361:20053511}

\bibitem[{Ehrenreich {et~al.}(2020)Ehrenreich, Lovis, Allart, Osorio, Pepe,
  Cristiani, Rebolo, Santos, Borsa, Demangeon, Dumusque, Hernández,
  Casasayas-Barris, Ségransan, Sousa, Abreu, Adibekyan, Affolter, Prieto, \&
  Zerbi}]{Ehrenreich2020}
Ehrenreich, D., Lovis, C., Allart, R., {et~al.} 2020, Nature, 1,
  \dodoi{10.1038/s41586-020-2107-1}

\bibitem[{{Esteves} {et~al.}(2017){Esteves}, {de Mooij}, {Jayawardhana},
  {Watson}, \& {de Kok}}]{Esteves2017}
{Esteves}, L.~J., {de Mooij}, E. J.~W., {Jayawardhana}, R., {Watson}, C., \&
  {de Kok}, R. 2017, \aj, 153, 268, \dodoi{10.3847/1538-3881/aa7133}

\bibitem[{{Evans} {et~al.}(2017){Evans}, {Sing}, {Kataria}, {Goyal}, {Nikolov},
  {Wakeford}, {Deming}, {Marley}, {Amundsen}, {Ballester}, {Barstow},
  {Ben-Jaffel}, {Bourrier}, {Buchhave}, {Cohen}, {Ehrenreich}, {Garc{\'\i}a
  Mu{\~n}oz}, {Henry}, {Knutson}, {Lavvas}, {Lecavelier Des Etangs}, {Lewis},
  {L{\'o}pez-Morales}, {Mandell}, {Sanz-Forcada}, {Tremblin}, \&
  {Lupu}}]{Evans2017}
{Evans}, T.~M., {Sing}, D.~K., {Kataria}, T., {et~al.} 2017, \nat, 548, 58,
  \dodoi{10.1038/nature23266}

\bibitem[{{Fortney} {et~al.}(2008){Fortney}, {Lodders}, {Marley}, \&
  {Freedman}}]{Fortney2008}
{Fortney}, J.~J., {Lodders}, K., {Marley}, M.~S., \& {Freedman}, R.~S. 2008,
  \apj, 678, 1419, \dodoi{10.1086/528370}

\bibitem[{{Gibson} {et~al.}(2020){Gibson}, {Merritt}, {Nugroho}, {Cubillos},
  {de Mooij}, {Mikal-Evans}, {Fossati}, {Lothringer}, {Nikolov}, {Sing},
  {Spake}, {Watson}, \& {Wilson}}]{Gibson2020}
{Gibson}, N.~P., {Merritt}, S., {Nugroho}, S.~K., {et~al.} 2020, \mnras, 493,
  2215, \dodoi{10.1093/mnras/staa228}

\bibitem[{{Grimm} \& {Heng}(2015)}]{Grimm_Heng2015}
{Grimm}, S.~L., \& {Heng}, K. 2015, \apj, 808, 182,
  \dodoi{10.1088/0004-637X/808/2/182}

\bibitem[{{Hawker} {et~al.}(2018){Hawker}, {Madhusudhan}, {Cabot}, \&
  {Gandhi}}]{Hawker2018}
{Hawker}, G.~A., {Madhusudhan}, N., {Cabot}, S. H.~C., \& {Gandhi}, S. 2018,
  \apjl, 863, L11, \dodoi{10.3847/2041-8213/aac49d}

\bibitem[{{H{\'e}brard} {et~al.}(2013){H{\'e}brard}, {Collier Cameron},
  {Brown}, {D{\'\i}az}, {Faedi}, {Smalley}, {Anderson}, {Armstrong}, {Barros},
  {Bento}, {Bouchy}, {Doyle}, {Enoch}, {G{\'o}mez Maqueo Chew}, {H{\'e}brard},
  {Hellier}, {Lendl}, {Lister}, {Maxted}, {McCormac}, {Moutou}, {Pollacco},
  {Queloz}, {Santerne}, {Skillen}, {Southworth}, {Tregloan-Reed}, {Triaud},
  {Udry}, {Vanhuysse}, {Watson}, {West}, \& {Wheatley}}]{Hebrard2013}
{H{\'e}brard}, G., {Collier Cameron}, A., {Brown}, D.~J.~A., {et~al.} 2013,
  \aap, 549, A134, \dodoi{10.1051/0004-6361/201220363}

\bibitem[{{Hoeijmakers} {et~al.}(2018){Hoeijmakers}, {Ehrenreich}, {Heng},
  {Kitzmann}, {Grimm}, {Allart}, {Deitrick}, {Wyttenbach}, {Oreshenko}, {Pino},
  {Rimmer}, {Molinari}, \& {Di Fabrizio}}]{Hoeij2018}
{Hoeijmakers}, H.~J., {Ehrenreich}, D., {Heng}, K., {et~al.} 2018, \nat, 560,
  453, \dodoi{10.1038/s41586-018-0401-y}

\bibitem[{{Hoeijmakers} {et~al.}(2019){Hoeijmakers}, {Ehrenreich}, {Kitzmann},
  {Allart}, {Grimm}, {Seidel}, {Wyttenbach}, {Pino}, {Nielsen}, {Fisher},
  {Rimmer}, {Bourrier}, {Cegla}, {Lavie}, {Lovis}, {Patzer}, {Stock}, {Pepe},
  \& {Heng}}]{Hoeij2019}
{Hoeijmakers}, H.~J., {Ehrenreich}, D., {Kitzmann}, D., {et~al.} 2019, \aap,
  627, A165, \dodoi{10.1051/0004-6361/201935089}

\bibitem[{{Ikoma} {et~al.}(2006){Ikoma}, {Guillot}, {Genda}, {Tanigawa}, \&
  {Ida}}]{Ikoma2006}
{Ikoma}, M., {Guillot}, T., {Genda}, H., {Tanigawa}, T., \& {Ida}, S. 2006,
  \apj, 650, 1150, \dodoi{10.1086/507088}

\bibitem[{{Kanagawa} {et~al.}(2018){Kanagawa}, {Muto}, {Okuzumi}, {Tanigawa},
  {Taki}, \& {Shibaike}}]{Kanagawa2018}
{Kanagawa}, K.~D., {Muto}, T., {Okuzumi}, S., {et~al.} 2018, \apj, 868, 48,
  \dodoi{10.3847/1538-4357/aae837}

\bibitem[{{Kataria} {et~al.}(2016){Kataria}, {Sing}, {Lewis}, {Visscher},
  {Showman}, {Fortney}, \& {Marley}}]{Kataria2016}
{Kataria}, T., {Sing}, D.~K., {Lewis}, N.~K., {et~al.} 2016, \apj, 821, 9,
  \dodoi{10.3847/0004-637X/821/1/9}

\bibitem[{{Kitzmann} {et~al.}(2018){Kitzmann}, {Heng}, {Rimmer}, {Hoeijmakers},
  {Tsai}, {Malik}, {Lendl}, {Deitrick}, \& {Demory}}]{Kitzmann2018}
{Kitzmann}, D., {Heng}, K., {Rimmer}, P.~B., {et~al.} 2018, \apj, 863, 183,
  \dodoi{10.3847/1538-4357/aace5a}

\bibitem[{{Knutson} {et~al.}(2010){Knutson}, {Howard}, \&
  {Isaacson}}]{Knutson2010}
{Knutson}, H.~A., {Howard}, A.~W., \& {Isaacson}, H. 2010, \apj, 720, 1569,
  \dodoi{10.1088/0004-637X/720/2/1569}

\bibitem[{{Kreidberg}(2015)}]{Laura2015}
{Kreidberg}, L. 2015, \pasp, 127, 1161, \dodoi{10.1086/683602}

\bibitem[{{Kurucz}(2018)}]{Kurucz2018}
{Kurucz}, R.~L. 2018, Astronomical Society of the Pacific Conference Series,
  Vol. 515, {Including All the Lines: Data Releases for Spectra and Opacities
  through 2017}, 47

\bibitem[{{Line} {et~al.}(2014){Line}, {Knutson}, {Wolf}, \& {Yung}}]{Line2014}
{Line}, M.~R., {Knutson}, H., {Wolf}, A.~S., \& {Yung}, Y.~L. 2014, \apj, 783,
  70, \dodoi{10.1088/0004-637X/783/2/70}

\bibitem[{{Louden} \& {Wheatley}(2015)}]{Louden_Wheatley2015}
{Louden}, T., \& {Wheatley}, P.~J. 2015, \apjl, 814, L24,
  \dodoi{10.1088/2041-8205/814/2/L24}

\bibitem[{{Madhusudhan} {et~al.}(2014){Madhusudhan}, {Amin}, \&
  {Kennedy}}]{Madhusudhan2014}
{Madhusudhan}, N., {Amin}, M.~A., \& {Kennedy}, G.~M. 2014, \apjl, 794, L12,
  \dodoi{10.1088/2041-8205/794/1/L12}

\bibitem[{{Madhusudhan} \& {Seager}(2009)}]{Mad_Sea2009}
{Madhusudhan}, N., \& {Seager}, S. 2009, \apj, 707, 24,
  \dodoi{10.1088/0004-637X/707/1/24}

\bibitem[{{Mayor} \& {Queloz}(1995)}]{MayorQueloz1995}
{Mayor}, M., \& {Queloz}, D. 1995, \nat, 378, 355, \dodoi{10.1038/378355a0}

\bibitem[{{Mazeh} {et~al.}(2007){Mazeh}, {Tamuz}, \& {Zucker}}]{Mazeh2007}
{Mazeh}, T., {Tamuz}, O., \& {Zucker}, S. 2007, Astronomical Society of the
  Pacific Conference Series, Vol. 366, {The Sys-Rem Detrending Algorithm:
  Implementation and Testing}, ed. C.~{Afonso}, D.~{Weldrake}, \& T.~{Henning},
  119

\bibitem[{Mbarek \& Kempton(2016)}]{Mbarek2016}
Mbarek, R., \& Kempton, E. M.-R. 2016, The Astrophysical Journal, 827, 121,
  \dodoi{10.3847/0004-637x/827/2/121}

\bibitem[{McKemmish {et~al.}(2019)McKemmish, Masseron, Hoeijmakers,
  Pérez-Mesa, Grimm, Yurchenko, \& Tennyson}]{TiOtoto}
McKemmish, L.~K., Masseron, T., Hoeijmakers, H.~J., {et~al.} 2019, Monthly
  Notices of the Royal Astronomical Society, 488, 2836,
  \dodoi{10.1093/mnras/stz1818}

\bibitem[{{Merritt} {et~al.}(2020){Merritt}, {Gibson}, {Nugroho}, {de Mooij},
  {Hooton}, {Matthews}, {McKemmish}, {Mikal-Evans}, {Nikolov}, {Sing}, {Spake},
  \& {Watson}}]{Merritt2020}
{Merritt}, S.~R., {Gibson}, N.~P., {Nugroho}, S.~K., {et~al.} 2020, arXiv
  e-prints, arXiv:2002.02795.
\newblock \doarXiv{2002.02795}

\bibitem[{{Narita} {et~al.}(2005){Narita}, {Suto}, {Winn}, {Turner}, {Aoki},
  {Leigh}, {Sato}, {Tamura}, \& {Yamada}}]{Narita2005}
{Narita}, N., {Suto}, Y., {Winn}, J.~N., {et~al.} 2005, \pasj, 57, 471,
  \dodoi{10.1093/pasj/57.3.471}

\bibitem[{{Noguchi} {et~al.}(2002){Noguchi}, {Aoki}, {Kawanomoto}, {Ando},
  {Honda}, {Izumiura}, {Kambe}, {Okita}, {Sadakane}, {Sato}, {Tajitsu},
  {Takada-Hidai}, {Tanaka}, {Watanabe}, \& {Yoshida}}]{Noguchi2002}
{Noguchi}, K., {Aoki}, W., {Kawanomoto}, S., {et~al.} 2002, \pasj, 54, 855,
  \dodoi{10.1093/pasj/54.6.855}

\bibitem[{{Nugroho} {et~al.}(2020{\natexlab{a}}){Nugroho}, {Gibson}, {de
  Mooij}, {Herman}, {Watson}, {Kawahara}, \& {Merritt}}]{Nugroho2020_wp33fe}
{Nugroho}, S.~K., {Gibson}, N.~P., {de Mooij}, E. J.~W., {et~al.}
  2020{\natexlab{a}}, \apjl, 898, L31, \dodoi{10.3847/2041-8213/aba4b6}

\bibitem[{{Nugroho} {et~al.}(2020{\natexlab{b}}){Nugroho}, {Gibson}, {de
  Mooij}, {Watson}, {Kawahara}, \& {Merritt}}]{Nugroho2020}
---. 2020{\natexlab{b}}, arXiv e-prints, arXiv:2003.04856.
\newblock \doarXiv{2003.04856}

\bibitem[{{Nugroho} {et~al.}(2017){Nugroho}, {Kawahara}, {Masuda}, {Hirano},
  {Kotani}, \& {Tajitsu}}]{Nugroho2017}
{Nugroho}, S.~K., {Kawahara}, H., {Masuda}, K., {et~al.} 2017, \aj, 154, 221,
  \dodoi{10.3847/1538-3881/aa9433}

\bibitem[{{{\"O}berg} {et~al.}(2011){{\"O}berg}, {Murray-Clay}, \&
  {Bergin}}]{Oberg2011}
{{\"O}berg}, K.~I., {Murray-Clay}, R., \& {Bergin}, E.~A. 2011, \apjl, 743,
  L16, \dodoi{10.1088/2041-8205/743/1/L16}

\bibitem[{{Parmentier} {et~al.}(2013){Parmentier}, {Showman}, \&
  {Lian}}]{Parmentier2013}
{Parmentier}, V., {Showman}, A.~P., \& {Lian}, Y. 2013, \aap, 558, A91,
  \dodoi{10.1051/0004-6361/201321132}

\bibitem[{{Parmentier} {et~al.}(2018){Parmentier}, {Line}, {Bean}, {Mansfield},
  {Kreidberg}, {Lupu}, {Visscher}, {D{\'e}sert}, {Fortney}, {Deleuil},
  {Arcangeli}, {Showman}, \& {Marley}}]{Parmentier2018}
{Parmentier}, V., {Line}, M.~R., {Bean}, J.~L., {et~al.} 2018, \aap, 617, A110,
  \dodoi{10.1051/0004-6361/201833059}

\bibitem[{{Plez}(1998)}]{Plez1998}
{Plez}, B. 1998, \aap, 337, 495

\bibitem[{{Pont} {et~al.}(2009){Pont}, {Gilliland}, {Knutson}, {Holman}, \&
  {Charbonneau}}]{Pont2009}
{Pont}, F., {Gilliland}, R.~L., {Knutson}, H., {Holman}, M., \& {Charbonneau},
  D. 2009, \mnras, 393, L6, \dodoi{10.1111/j.1745-3933.2008.00582.x}

\bibitem[{{Rasio} \& {Ford}(1996)}]{Rasio1996}
{Rasio}, F.~A., \& {Ford}, E.~B. 1996, Science, 274, 954,
  \dodoi{10.1126/science.274.5289.954}

\bibitem[{{S{\'a}nchez-L{\'o}pez} {et~al.}(2019){S{\'a}nchez-L{\'o}pez},
  {Alonso-Floriano}, {L{\'o}pez-Puertas}, {Snellen}, {Funke}, {Nagel}, {Bauer},
  {Amado}, {Caballero}, {Czesla}, {Nortmann}, {Pall{\'e}}, {Salz}, {Reiners},
  {Ribas}, {Quirrenbach}, {Anglada-Escud{\'e}}, {B{\'e}jar},
  {Casasayas-Barris}, {Galad{\'\i}-Enr{\'\i}quez}, {Guenther}, {Henning},
  {Kaminski}, {K{\"u}rster}, {Lamp{\'o}n}, {Lara}, {Montes}, {Morales},
  {Stangret}, {Tal-Or}, {Sanz-Forcada}, {Schmitt}, {Zapatero Osorio}, \&
  {Zechmeister}}]{Sanchez2019}
{S{\'a}nchez-L{\'o}pez}, A., {Alonso-Floriano}, F.~J., {L{\'o}pez-Puertas}, M.,
  {et~al.} 2019, \aap, 630, A53, \dodoi{10.1051/0004-6361/201936084}

\bibitem[{{Sato} {et~al.}(2005){Sato}, {Fischer}, {Henry}, {Laughlin},
  {Butler}, {Marcy}, {Vogt}, {Bodenheimer}, {Ida}, {Toyota}, {Wolf}, {Valenti},
  {Boyd}, {Johnson}, {Wright}, {Ammons}, {Robinson}, {Strader}, {McCarthy},
  {Tah}, \& {Minniti}}]{Sato2005}
{Sato}, B., {Fischer}, D.~A., {Henry}, G.~W., {et~al.} 2005, \apj, 633, 465,
  \dodoi{10.1086/449306}

\bibitem[{{Schreier} {et~al.}(2019){Schreier}, {Gimeno Garc{\'\i}a},
  {Hochstaffl}, \& {St{\"a}dt}}]{Schreier2019}
{Schreier}, F., {Gimeno Garc{\'\i}a}, S., {Hochstaffl}, P., \& {St{\"a}dt}, S.
  2019, Atmosphere, 10, 262, \dodoi{10.3390/atmos10050262}

\bibitem[{{Sing} {et~al.}(2019){Sing}, {Lavvas}, {Ballester}, {Lecavelier des
  Etangs}, {Marley}, {Nikolov}, {Ben-Jaffel}, {Bourrier}, {Buchhave}, {Deming},
  {Ehrenreich}, {Mikal-Evans}, {Kataria}, {Lewis}, {L{\'o}pez-Morales},
  {Garc{\'\i}a Mu{\~n}oz}, {Henry}, {Sanz-Forcada}, {Spake}, {Wakeford}, \&
  {PanCET Collaboration}}]{Sing2019}
{Sing}, D.~K., {Lavvas}, P., {Ballester}, G.~E., {et~al.} 2019, \aj, 158, 91,
  \dodoi{10.3847/1538-3881/ab2986}

\bibitem[{{Snellen} {et~al.}(2010){Snellen}, {de Kok}, {de Mooij}, \&
  {Albrecht}}]{Snellen2010}
{Snellen}, I. A.~G., {de Kok}, R.~J., {de Mooij}, E. J.~W., \& {Albrecht}, S.
  2010, \nat, 465, 1049, \dodoi{10.1038/nature09111}

\bibitem[{{Spiegel} {et~al.}(2009){Spiegel}, {Silverio}, \&
  {Burrows}}]{Spiegel2009}
{Spiegel}, D.~S., {Silverio}, K., \& {Burrows}, A. 2009, \apj, 699, 1487,
  \dodoi{10.1088/0004-637X/699/2/1487}

\bibitem[{{Stassun} {et~al.}(2017){Stassun}, {Collins}, \&
  {Gaudi}}]{Stassun2017}
{Stassun}, K.~G., {Collins}, K.~A., \& {Gaudi}, B.~S. 2017, \aj, 153, 136,
  \dodoi{10.3847/1538-3881/aa5df3}

\bibitem[{{Stock} {et~al.}(2018){Stock}, {Kitzmann}, {Patzer}, \&
  {Sedlmayr}}]{Stock2018}
{Stock}, J.~W., {Kitzmann}, D., {Patzer}, A. B.~C., \& {Sedlmayr}, E. 2018,
  \mnras, 479, 865, \dodoi{10.1093/mnras/sty1531}

\bibitem[{{Tajitsu} {et~al.}(2010){Tajitsu}, {Aoki}, {Kawanomoto}, \&
  {Narita}}]{Tajitsu2010}
{Tajitsu}, A., {Aoki}, W., {Kawanomoto}, S., \& {Narita}, N. 2010, Publications
  of the National Astronomical Observatory of Japan, 13, 1

\bibitem[{{Tamuz} {et~al.}(2005){Tamuz}, {Mazeh}, \& {Zucker}}]{Tamuz2005}
{Tamuz}, O., {Mazeh}, T., \& {Zucker}, S. 2005, \mnras, 356, 1466,
  \dodoi{10.1111/j.1365-2966.2004.08585.x}

\bibitem[{{Tanaka} {et~al.}(2002){Tanaka}, {Takeuchi}, \& {Ward}}]{Tanaka2002}
{Tanaka}, H., {Takeuchi}, T., \& {Ward}, W.~R. 2002, \apj, 565, 1257,
  \dodoi{10.1086/324713}

\bibitem[{{Turner} {et~al.}(2020){Turner}, {de Mooij}, {Jayawardhana}, {Young},
  {Fossati}, {Koskinen}, {Lothringer}, {Karjalainen}, \&
  {Karjalainen}}]{Turner2020}
{Turner}, J.~D., {de Mooij}, E. J.~W., {Jayawardhana}, R., {et~al.} 2020,
  \apjl, 888, L13, \dodoi{10.3847/2041-8213/ab60a9}

\bibitem[{{Valenti} {et~al.}(1995){Valenti}, {Butler}, \&
  {Marcy}}]{Valenti1995}
{Valenti}, J.~A., {Butler}, R.~P., \& {Marcy}, G.~W. 1995, \pasp, 107, 966,
  \dodoi{10.1086/133645}

\bibitem[{{Wakeford} \& {Sing}(2015)}]{WakeSing2015}
{Wakeford}, H.~R., \& {Sing}, D.~K. 2015, \aap, 573, A122,
  \dodoi{10.1051/0004-6361/201424207}

\bibitem[{{Winn} {et~al.}(2004){Winn}, {Suto}, {Turner}, {Narita}, {Frye},
  {Aoki}, {Sato}, \& {Yamada}}]{Winn2004}
{Winn}, J.~N., {Suto}, Y., {Turner}, E.~L., {et~al.} 2004, \pasj, 56, 655,
  \dodoi{10.1093/pasj/56.4.655}

\bibitem[{{Yan} {et~al.}(2019){Yan}, {Casasayas-Barris}, {Molaverdikhani},
  {Alonso-Floriano}, {Reiners}, {Pall{\'e}}, {Henning}, {Molli{\`e}re}, {Chen},
  {Nortmann}, {Snellen}, {Ribas}, {Quirrenbach}, {Caballero}, {Amado},
  {Azzaro}, {Bauer}, {Cort{\'e}s Contreras}, {Czesla}, {Khalafinejad}, {Lara},
  {L{\'o}pez-Puertas}, {Montes}, {Nagel}, {Oshagh}, {S{\'a}nchez-L{\'o}pez},
  {Stangret}, \& {Zechmeister}}]{Yan2019}
{Yan}, F., {Casasayas-Barris}, N., {Molaverdikhani}, K., {et~al.} 2019, \aap,
  632, A69, \dodoi{10.1051/0004-6361/201936396}

\end{thebibliography}

\end{document}